THE UNIVERSITY OF MELBOURNE

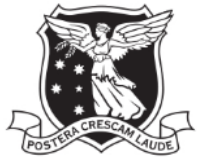

Learning to Rank with Small Set of Ground Truth Data

Jiashu Wu

Submitted in partial fulfilment of the requirements for the degree of

MASTER OF INFORMATION TECHNOLOGY

OCTOBER 2019

SCHOOL OF COMPUTING AND INFORMATION SYSTEMS

MELBOURNE SCHOOL OF ENGINEERING

THE UNIVERSITY OF MELBOURNE

# Declaration

I certify that:

This thesis does not incorporate without acknowledgement any material previously submitted for a degree or diploma in any university, and that to the best of my knowledge and belief it does not contain any material previously published or written by another person where due reference is not made in the text.

The thesis is 11235 words in length (excluding text in images, tables and references).

*Jiashu Wu*

Oct 2019

# Acknowledgement

I would like to thank my family and my supervisor, Professor Rui Zhang, for providing guidance, support, and feedback throughout this research project.

# Table of contents

# List of figures

# List of tables

# 1. Abstract

Over the past decades, researchers had put lots of effort investigating ranking techniques used to rank query results retrieved during information retrieval, or to rank the recommended products in recommender systems. In this project, we aim to investigate searching, ranking, as well as recommendation techniques to help to realize a university academia searching platform. Unlike the usual information retrieval scenarios where lots of ground truth ranking data is present, in our case, we have only limited ground truth knowledge regarding the academia ranking. For instance, given some search queries, we only know a few researchers who are highly relevant and thus should be ranked at the top, and for some other search queries, we have no knowledge about which researcher should be ranked at the top at all. The limited amount of ground truth data makes some of the conventional ranking techniques and evaluation metrics become infeasible, and this is a huge challenge we faced during this project.

Two versions of approaches have been implemented and thoroughly experimented. One of them utilizes BM25F, matrix decomposition algorithms like Latent Semantic Analysis (LSA), Non-negative Matrix Factorization (NMF), as well as Factorization Machines (FM) models such as Neural Factorization Machines (NFM). Another version of approach combines BM25F information, as well as confidence score calculated and deployed by Microsoft Academic Graph (MAG) knowledge base. After thorough comparisons and observations, the result produced by the second approach outperforms the result yielded by the first approach, and hence eventually, the second approach is currently being deployed to our searching platform at the end.

Besides searching, other functionalities have also been researched, experimented and built for this platform, for instance, browsing researchers by a concept hierarchy tree, which also involves ranking techniques in it, as well as the query auto-complete functionality, which can assist users to auto-complete their search queries.

Overall, this project enhances the user's academia searching experience to a large extent, it helps to achieve an academic searching platform which includes researchers, publications and fields of study information, which will be beneficial not only to the university faculties but also to students' research experiences.

# 2. Literature reviews

Initially, several survey papers on searching and recommendation have been studied. [1] summarizes methods that can be used to tackle matching problems in searching and recommendation. The matching problem has been a key problem in searching and recommendation to measure the relevance between documents and search query or measure the degree of matching between a user's interest and candidate items. Many traditional algorithms and techniques have been used to address the problem, the relevance measures like TF-IDF and BM25 can effectively measure the relatedness needed for searching and recommendation, latent space matching algorithms like Partial Least Square and Regularized Mapping to Latent Space, have been exploited to bridging the semantic gaps, one of the biggest challenges faced by both searching and recommendation. On the other hand, thanks to the emergence of deep learning, now, algorithms like Neural Matrix Factorization (NMF) can be utilized to solve the matching and ranking problems in searching and recommendation effectively. What's more, [2] suggested that knowledge bases can also be combined with algorithms to perform searching and recommendation in a hybrid way. Knowledge from knowledge bases can help tremendously during searching and recommendation and outperforms some of the algorithms that do not take this knowledge into account.

Hence, in this project, various kinds of techniques and algorithms have been investigated, and different versions of solutions have been implemented and experimented. Given that these techniques do not necessarily overlap with each other, therefore, the literature reviews are done separately and therefore will be presented in the form of several subsections.

## 2.1 Conventional machine learning approaches

Conventional machine learning algorithms have been applied in searching and recommendation for a long time. [3] demonstrates that matrix factorization algorithms can characterize both users and items by vectors of factors, which will lead to successful recommendations. The method they proposed took the top spot in the Netflix prize competition. [4] proposed and utilized Factorization Machines (FM) to provide rating predictions. This method is computationally efficient, as the computation takes linear time in the number of both context variables and the factorization size. Empirically the Factorization Machines produce a satisfactory result in both prediction quality and run time. [5] proposed Factored Item Similarity Models (FISM), which could be used for top-N recommendation, and it can generate high-quality recommendations even on very sparse datasets. [6] proposed SVD++, which fuses user-based and item-based collaborative filtering, this approach dramatically improves the recommendation quality.

## 2.2 Deep learning-related approaches

With the fast emergence of deep learning techniques and its application, deep learning-based methods have been applied to the field of recommendation. [7] summarizes the usage of deep learning techniques in modern recommendation systems, thanks to the power of deep learning, various recommendation techniques have been significantly advanced. [8] proposed neural network-based collaborative filtering, which is a generic method that can generalize to different data and model the user-item interactions with non-linearities. Extensive experiments on datasets such as MovieLens shows a state-of-the-art recommendation performance. [9] proposed a novel matrix factorization model with neural network architecture. This neural network architecture can learn a low dimensional latent space and then represent the user and item features in this latent space. Experimental results demonstrate a promising recommendation performance. [10] proposed Wide and Deep learning, which combines wide linear models and deep neural networks. This method can tackle the case in which user-item interactions are sparse and high-rank, hence this method is very generalizable. The experiment performed on the Google play store demonstrates a significant increase in app acquisition due to better recommendations. [11] proposed Neural Factorization Machines (NFM), which combines both the linearity of Factorization Machine, as well as the non-linearity introduced by the neural networks to represent higher-order feature interactions. Experimental evidence on Frappe mobile app usage log dataset and MovieLens dataset indicate that the newly proposed Neural Factorization Machines model is more expressive than Factorization Machines, which does not have hidden layers.

## 2.3 Knowledge base related approaches

Knowledge base related techniques are also popular ways to generate recommendations, it outperforms the traditional recommendation approaches such as collaborative filtering in several ways. [12] experimented with integrating a knowledge base of product semantics to provide recommendations on an online retailer system. By understanding the features of the products, it can recommend products in the same class as those products that the user has purchased just as traditional collaborative filtering can do. What makes it stands out is that it can also learn user's preferences using the knowledge it has regarding the products, then recommend products that are likely to match the user's taste, which demonstrates the advantages of this hybrid approach. [13] demonstrates the combination of collaborative filtering with domain knowledge about the products is a very powerful recommendation technique. This hybrid approach is also applied in real applications such as recommending restaurants to users. In this case, similar restaurants are found by examining the similarity of features between restaurants, and these important features come from a restaurant knowledge base. [14] applied knowledge-based recommender system to recommend travel destinations to users. By integrating a knowledge base that contains airfares, hotel and vacation packages information, then generate recommendations base on the relevancy between user's preferences and the destinations domain knowledge information. This turns out to be a very effective approach, as it produces a nearly fivefold increase in the product selling rate. What's more, the approach they used to build the knowledge base is a semi-automatic process, user's feedback is later used to enhance the domain knowledge in the knowledge base, and hence the enhanced knowledge can further yield a better recommendation. By using a knowledge base that gradually keeping track of user's tastes and market changes, it outperforms traditional knowledge bases that operate in a static manner.

### 2.3.1 Microsoft Academic Graph

Among the literature we reviewed, the knowledge base that is most relevant to this project is the Microsoft Academic Graph (MAG) knowledge base, introduced by [15]. As the largest publicly available academic knowledge base to-date, it contains handfuls of useful knowledge regarding publications, authors, institutions, field-of-studies (FOS), etc. There are several properties of this knowledge base that attract our attention and make it a useful component to be used in our project:

1) Its concept discovery mechanism mines academic concepts from various domains, with reference to existing classification, corresponding Wikipedia information, and also with the help of human inspections to improve the quality.
2) Semantic closeness based on graph link analysis is computed and entity type based filtering and enrichment is applied to filter noises and eliminate non-academic words, hence the FOS identified is highly refined with high quality.
3) Publications written by billions of authors are tagged with certain confidence scores using the FOS identified during concept discovery. These tags become very important features of these publications, as they characterize the topic, content and main focus of each publication.
4) By aggregating the FOS tags of publications of each author, it becomes features that depict the research interest and focus of an author.
5) A concept hierarchy is built, which includes crucial information like parent FOS, child FOS, related FOS. This clearly demonstrates the hierarchical relationships and parallel relationships between FOSs.

The diagram below demonstrates some example FOS that are contained in the Microsoft Academic Graph knowledge base, and their corresponding FOS relationships.

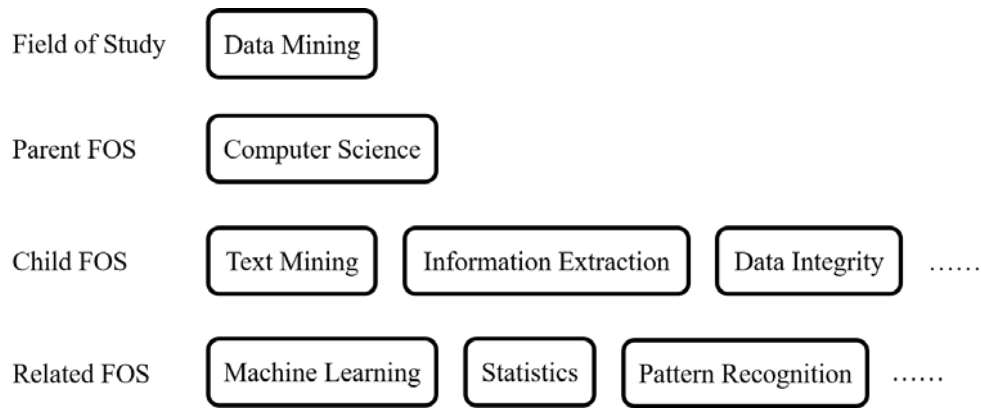

Figure 1. Demonstration of examples of Field of Studies (FOS)

These excellent functionalities and properties of MAG make it a very worthwhile candidate to explore and experiment in our project.

## 2.4 Information retrieval functions

Huge amounts of prior researches demonstrate that the Term Frequency Inverse Document Frequency (TFIDF) is an effective measure to determine the relevancy between search query and documents. [16] examines the result of applying TFIDF to match words in queries to match documents. It is a straightforward way of measuring the degree of relevancy, hence, it forms the basis of many query retrieval systems.

Besides TFIDF, BM25 is also one of the most successful text retrieval algorithms. In recent years, [17] proposed to use BM25 for semantic search, and [18] applied BM25 to retrieve patents that are relevant to a given patent.

One of the variations of BM25 is BM25F, introduced by [19], the difference between BM25F and BM25 is that BM25F considers fields in a given article differently, it integrates content features across multiple fields in each document, regardless of the length of fields, for instance, it still works well even if the title field and abstract field of an article have very different length. It turns out that by integrating different field features for a document, it produces better results in TREC Web track. The details of BM25F are as follows:

$$\bar{x}_{d,f,t} := \frac{x_{d,f,t}}{(1 + B_f(\frac{l_{d,f}}{l_f} - 1))}$$

$$\bar{x}_{d,t} = \sum_f W_f \cdot \bar{x}_{d,f,t}$$

$$BM25F(d) := \sum_{t \in q \cap d} \frac{\bar{x}_{d,t}}{K1 + \bar{x}_{d,t}} w_t$$

where $f$ indicates fields like TITLE, BODY, etc. $B_f$ is a field-dependent normalizing parameter. $l_{d,f}$ indicates the length of that field, $l_f$ is the average field length of that field type. $x_{d,f,t}$ is the term frequency of term $t$ in the field type $f$ of document $d$. $W_f$ is the weight we assigned for each different field in a document, for instance, if we assign $W_{title}$ to 1.2 and $W_{body}$ to 1.0, then this literally indicates that we think the title should be treated more importantly than the body of the document. $w_t$ is the RSJ relevance weight for term $t$, or the IDF weight if such information is not available. $K_1$ is a saturating parameter. There also exists researches that suggest general parameter ranges based on empirical observations and studies, which make the hyperparameter tuning in our project easier.

# 3. Methodology

This project contains several parts to research, implement and experiment. The core functionalities we researched and implemented for the academic searching platform are as follows:

1) Searching: user enters a search query, then relevant researchers will be ranked and displayed in descending order of relevancy.
2) Browsing: Researchers are categorised using a FOS concept hierarchy tree. Under each tree leave, i.e., a FOS term, relevant researches are ranked in descending order of relevancy.
3) Term extraction, which extracts texts from publications, user search queries, etc. into several recognisable terms that can be found in a term dictionary.
4) Query auto-completion, which is a functionality complements the searching service. This functionality can help users to complete the search query by suggesting popular search query terms.
5) To ship the deployment, a stable TCP server and client are developed to make the services described above deployable.

The methodologies, algorithms and design details will be explained in sections below.

## 3.1 Searching

Inspired from the insights we get from literature reviews, we decide to implement two versions of solutions to do the searching, utilizing two different kinds of approaches. The first version involves BM25F scores, a handcrafted matrix transformation formula, matrix decompositions algorithms, deep learning-based matrix factorization methods. The second version utilizes BM25F scores and MAG knowledge base. I will demonstrate the methodology details of both versions of solutions in the following subsections.

### 3.1.1 Searching using matrix factorization algorithms

The diagram below is a general overview of the procedures we take when searching using matrix factorization algorithms. Note that the process will only be done once, then the results will be dumped for future use.

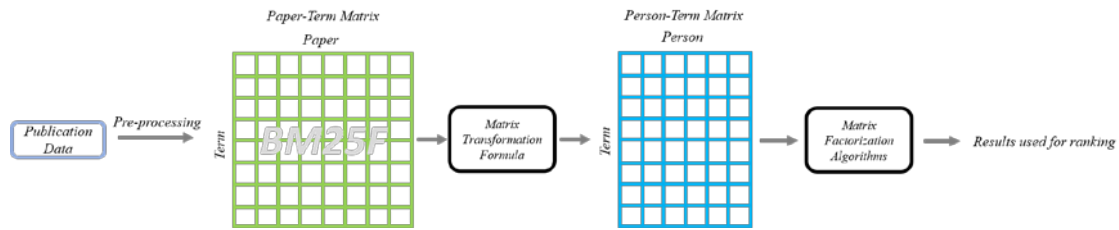

Figure 2. Procedures for searching using matrix factorization algorithms

### 3.1.1.1 Data source used

The data we used in this implementation including the following: a publication database downloaded from University of Melbourne FindExperts platform, the database contains useful information such as publication's title, publication's abstract, the name of the journal or conference that the publication appears in, as well as some keywords describing the publication. Besides the publication database, we also downloaded Wikipedia pages-articles dump database, which contains 19549637 terms, while after cleansing and pre-processing such as eliminating noisy entries, there remain 10048143 terms which will be used in our project.

### 3.1.1.2 Publication data pre-processing

Firstly, publication title, publication abstract, publication keywords, and publication journal and conference name in publication data are pre-processed via the following steps: lowercasing, remove unwanted copyright information like "© Copyright 2019. All rights reserved", as this information doesn't help in our case at all. Duplicate punctuations and duplicate white spaces are also cleaned. After that, we performed sentence segmentation to segment paragraphs into sentences. Finally, we performed the term extraction to extract terms in publications that belong to the term dictionary. The details of term extraction mechanism will be explained in section 3.3 of this report.

The figure below illustrates an example of pre-processing procedures and corresponding outputs.

Automatic paraphrasing is an important component in many natural language processing tasks. In this article we present a new parallel corpus with paraphrase annotation. © 2008 Association for Computational Linguistics.

↓ Lowercasing
  Removing Copyright
  Removing Duplicates

automatic paraphrasing is an important component in many natural language processing tasks. in this article we present a new parallel corpus with paraphrase annotation.

↓ Sentence Segmentation

automatic paraphrasing is an important component in many natural language processing tasks.
in this article we present a new parallel corpus with paraphrase annotation.

↓ Term Extraction

automatic paraphrasing, important, component, natural language processing, task, article, present, parallel corpus, paraphrase annotation

Figure 3. An illustrative example of pre-processing procedures and results

After pre-processing, the extracted terms will be used to construct the paper-term matrix.

### 3.1.1.3 Paper-term matrix construction

After finishing the pre-processing procedures, we construct the paper-term matrix. The columns of this matrix indicate each publication we have in our database, while the rows of this matrix indicate terms extracted from these publications. In each cell, it contains the BM25F score of this specific term in this specific publication.

For instance, the example above will have rows such as automatic paraphrasing, important, natural language processing, etc. stores the corresponding BM25F scores in this publication.

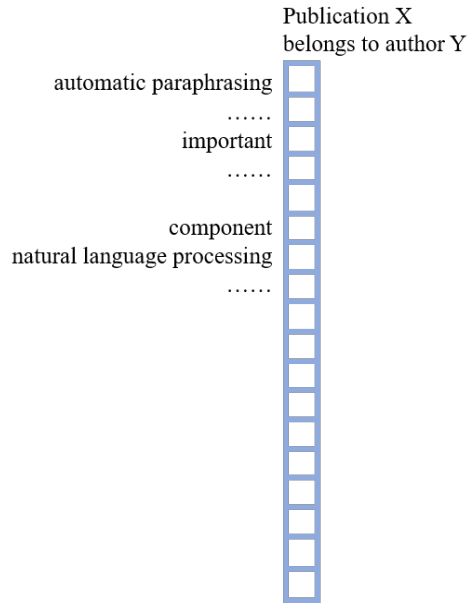

Figure 4. Illustration of paper-term matrix

Note that the figure above is used only to illustrate the idea of the paper-term matrix. In order to make the program space-efficient, the paper-term matrix is stored in the form of inverted index, rather than a huge matrix.

### 3.1.1.4 Usage of BM25F and its advantages

The usage of BM25F is inspired by our literature reviews. Since academic publications usually have title, abstract, main body, keywords, etc., the terms appear in the title should be treated with the highest weight, as the title is a very concise summary of the topic and content of the whole publication. The terms appear in the abstract section should also deserve a weight that is higher than the main body, as abstract is a compact summarization of the body content at a relatively high level. Terms extracted from journal and conference titles should also be assigned with a relatively high weight, since a publication will go through careful scrutiny by domain experts before being qualified to enter the journal and conference, hence, a publication which enters EMNLP is unlikely not relevant to natural language processing. Given that publications usually have these fields in it with varying importance weight, using BM25F is a suitable choice in this case, as BM25F is a variation of BM25 which treats fields differently. This huge advantage of BM25F makes it the most suitable measurement to be used in our project.

### 3.1.1.5 Matrices transformation

The second step is to transform the paper-term matrix we constructed in the previous step into the person-term matrix, using a matrix transformation formula we designed.

During our experimentations, three versions of formula have been tried.

### 3.1.1.5.1 Matrix transformation formula #1

Score of person of a term =
$$\frac{(\sum BM25F \text{ of this term from papers written by this person}+1)^{w1} * \log(\text{\# paper written by this person which contains this term}+1)^{w2}}{\log(\text{\# paper written by this person}+1)^{w3}}$$

This formula considers the sum of BM25F scores of this term from papers written by this person, the number of papers written by this person which contains this term, as well as the total number of papers written by this person. Note that the Add-one operations have been applied to prevent 0 values, and weight parameters w1 to w3 are used to make the formula tuneable. The intuition behind this formula is the following:

1) The person who has a higher sum of BM25F scores should end up with a higher score, as high BM25F score indicates high relevancy towards this term
2) The more paper is written by this person which contains this term, the more this person should be rewarded
3) The more paper a person wrote, the more likely he/she will include this term, hence, the formula is normalized by dividing the total number of paper written by this person

### 3.1.1.5.2 Issues faced by this formula

However, this formula faces the following problem:

Below are two example persons with the following term occurrence, respectively.

Example 1: person A has 10 papers in total, 5 of them contains the term "Natural Language Processing", each occurs 10 times, and the total sum of BM25F score is 20.

Example 2: person B has 10 papers in total, 5 of them contain the term "Natural Language Processing", and this term occurs 46 times in paper 1, and one time each in paper 2 to 5, the total sum of BM25F score is 20.

When this version of the formula is applied, then both of them will end up with exactly the same score. However, intuitively speaking, person A should be rewarded with a higher score, as he/she mentions this term evenly in his/her publications, and each publication contains this term relatively frequently. While person B only mentions this term once in 4 of his/her paper and only has one paper which contains this term frequently. The infrequent occurrences of this term in person B's papers may indicate that he/she is not expertized in this area, or his/her main research interest is not on this term. In practice, we do observe that some of the researchers from Faculty of Arts for instance applied "Artificial Intelligence" related techniques in one of his/her research paper, and the AI technique is only used as a helpful application in his/her art research, not as the main focus of the research paper and therefore not the major research interest of the researcher. Hence, this formula lacks discriminating power in this scenario.

### 3.1.1.5.3 Knowledge gained

The drawback faced by this formula tells us that the number of occurrences of a term solely does not guarantee that the research is relevant to this term. The distribution of the occurrences of this term should also be taken into account when designing the matrix transformation formula. If omitted, then the lack of discriminating power caused by this drawback will mistakenly let these application-oriented papers and researchers being considered as relevant, which is not what we expect. After taking this factor into account, it should be able to distinguish between those researchers who are truly expertized and those researchers who only use this term as an application or tool for his/her researchers in other areas, because those researchers who use this term as an application may mention this term in his/her paper quantitively, but not in-depth for the majority of his/her papers.

Hence, in order to tackle this issue, we redesigned our matrix transformation formula by taking term occurrence distribution into consideration.

### 3.1.1.5.4 Matrix transformation formula #2

Score of person of a term =
$$\frac{(\sum BM25F \text{ of this term from papers written by this person}+1)^{w1} * \log(\text{\# paper written by this person which contains this term}+1)^{w2}}{Sigmoid(Var(\text{\# of occurrence of this term in each paper written by this person contains this term}))^{w3} * \log(\text{\# paper written by this person}+1)^{w4}}$$

In order to tackle the challenge faced by our previous formula, we modified our design and experimented with this new formula version.

In the new formula we designed, a new variance term is introduced in the denominator of this new formula, the intuition behind this term is as follows:

1) In the example we mentioned above, person A's term occurrences (10, 10, 10, 10, 10) will have a variance of 0, hence it leads to a smaller denominator, hence the overall score becomes higher.
   While person B's term occurrences (46, 1, 1, 1, 1) will have a variance higher than 0, and hence it leads to an overall score lower than person A

2) Assume there is another person, person C, with term occurrences (1, 1, 1, 1, 1), this will also have a zero variance, however, the sum of BM25F in the nominator should be able to reward person A a higher score, and reward person C a lower score as this term occurs in person C's publications infrequently. The sum of BM25F should be able to distinguish these two persons when the variance is not distinguishing enough.

### 3.1.1.5.5 Matrix transformation formula #3
Besides formula #2, we also proposed another formula to tackle the problem.

Score of person of a term =
$$\frac{(\sum BM25F \text{ of this term from papers written by this person}+1)^{w1} * \log(\# \text{ paper written by this person which contains this term}+1)^{w2}}{(\tanh(\max(AVG_5 - avg_5, 0))+1)^{w3} * \log(\# \text{ paper written by this person}+1)^{w4}}$$

, where:

1) $AVG_5$: after finishing the calculation for the first time, we take top 5 ranked authors, take the top 5 papers from each person, $AVG_5$ is the average score of these 25 papers
2) $avg_5$: When calculating the score for a term, $avg_5$ stands for the average score of this term of this person's paper with the top 5 highest BM25F score
3) Before we begin the calculation for the first round, $AVG_5 = avg_5 = 0$

Instead of using the variance, we introduced a new term in the denominator of this formula. The intuition behind this formula is as follows:

1) When calculating the scores, we do it twice. For the first time of our calculation, we don't know who should be ranked in top yet, hence, in this case, $AVG_5 = avg_5 = 0$, without prior knowledge, this formula downgrades to formula #1. After using this formula to calculate the ranking results for the first time, we take the top 5 ranked persons, take top 5 papers from each person, calculate the average score of these 25 papers, denoted as $AVG_5$, then go back and use this formula to calculate the ranking again. Now, the $AVG_5$ becomes our prior knowledge, it tells us what kind of score should a person have to be considered as highly relevant. The second turn of calculation has the prior knowledge, and for a given person, if his top 5 paper's average score, denoted as $avg_5$, is above $AVG_5$, this indicates that he is more relevant than the top 5 persons in our previous round of calculation, hence, this will lead to a smaller denominator and higher overall score. If, on the other hand, this person's top 5 paper's average score is lower than $AVG5$, this indicates that this person is not as relevant as the gold standard, hence, this will lead to a larger denominator and a lower overall score. The first round of calculation may not be very precise, but it shouldn't be completely on the wrong track, and this new term in the denominator is basically comparing this given author with the "gold standard".

However, as we can see, these formulas all contain tunable hyperparameters, given that we only have limited ground truth knowledge about the ranking results, these formulas all face the challenge of hyperparameter tuning. We need to manually examine the ranking results of several different queries to tune the hyperparameter settings.

3.1.1.5.6 Thoughts regarding matrix transformation formula #3

Since matrix transformation formula #3 compares each researcher with the gold standard, i.e., highly ranked researchers in the first round of calculation, therefore, this formula basically involves a learning process, i.e., it is capable of decreases the score difference between highly relevant researchers and the gold standard, and can increase the score difference between irrelevant researchers and the gold standard. Under the scenario in which ground truth ranking data is limited, this formula may be able to perform learning to rank. The idea behind this formula is quite similar to the pseudo relevance feedback in information retrieval, instead of real user feedback, the top-ranked results are used as pseudo-user feedback to help improve the ranking. In our future work, it is possible for us to keep investigating this learning to rank technique to tackle the challenge of lacking ground truth ranking data.

### 3.1.1.6 Person-term matrix construction

The person-term matrix is a matrix whose columns represent each author, and rows represent each term. This matrix is generated by using the paper-term matrix as input, and then put into the matrix transformation formula we designed above.

Compared with the paper-term matrix, both matrices have exactly the same number of rows. The difference is in the number of columns, and the content of cells. The columns of the paper-term matrix are combined with regard to authors by the summation in the formula. If a researcher wrote 10 publications, then he will have 10 columns in the paper-term matrix, each column represents a publication written by him/her. After applying the summation in the matrix transformation formula, in the person-term matrix, this researcher will only occupy one column, which represents himself/herself. Also, cells contain the value calculated by the formula, rather than the raw BM25F scores.

Note that the person-term matrix is also stored in the form of inverted index in order to make the program becomes more space-efficient.

### 3.1.1.7 Matrix decomposition algorithms

After finish constructing the person-term matrix, we tried two matrix decomposition algorithms to decompose the matrix into latent vectors.

One of the algorithms we experimented with is the Latent Semantic Analysis (LSA), which decomposes the person-term matrix into the product of a term-latent matrix, a latent transformation matrix, and a latent-person matrix. Another algorithm we experimented is the Non-negative Matrix Factorization (NMF), the NMF decomposes the matrix into the product of two matrices, similar to the first and third matrix produced by LSA, but the NMF didn't generate the latent transformation matrix, which makes its result becomes more explainable.

After matrix decomposition, we rank documents in descending order of query vector – person vector cosine similarity. Note that we also attempted to normalize the BM25F scores in advance when we build the paper-term matrix so that the normalization in the denominator of cosine similarity can be omitted.

The following diagrams illustrate searching using LSA and NMF, respectively. Note that the pre-processing is also done before constructing the paper-term matrix, the diagram just omits the grey pre-processing box before the green matrix for simple illustration purposes.

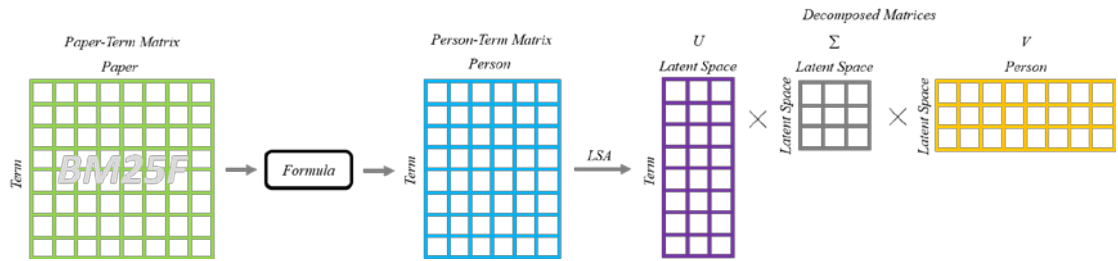

Figure 5. Procedures for searching using Latent Semantic Analysis

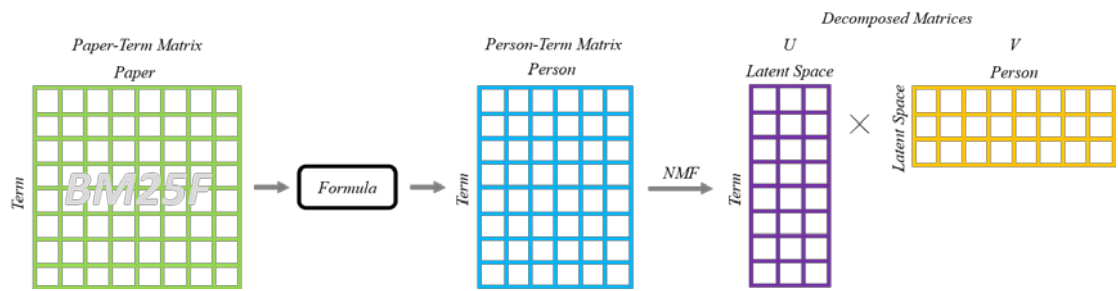

Figure 6. Procedures for searching using Non-negative Matrix Factorization

### 3.1.1.8 Neural Factorization Machines

We not only tried matrix decomposition algorithms, but also experimented deep learning-based algorithm, which is Neural Factorization Machines (NFM). According to the literature review, Neural Factorization Machines combines the linearity of Factorization Machines (FM) in modelling second-order feature interactions and the non-linearity introduced by deep learning models when modelling feature interactions with higher order. It also performs well under sparse settings, which suits our case. Empirical results demonstrate that the Neural Factorization Machines significantly outperform Factorization Machines, as well as other deep learning-based algorithms.

The diagram below illustrates the procedures of searching using NFM.

Note that in order to simplify the diagram, the pre-processing is not included in the diagram, also the Libfm file construction is omitted in the diagram. But they are all necessary procedures we actually take during our implementation of searching using NFM.

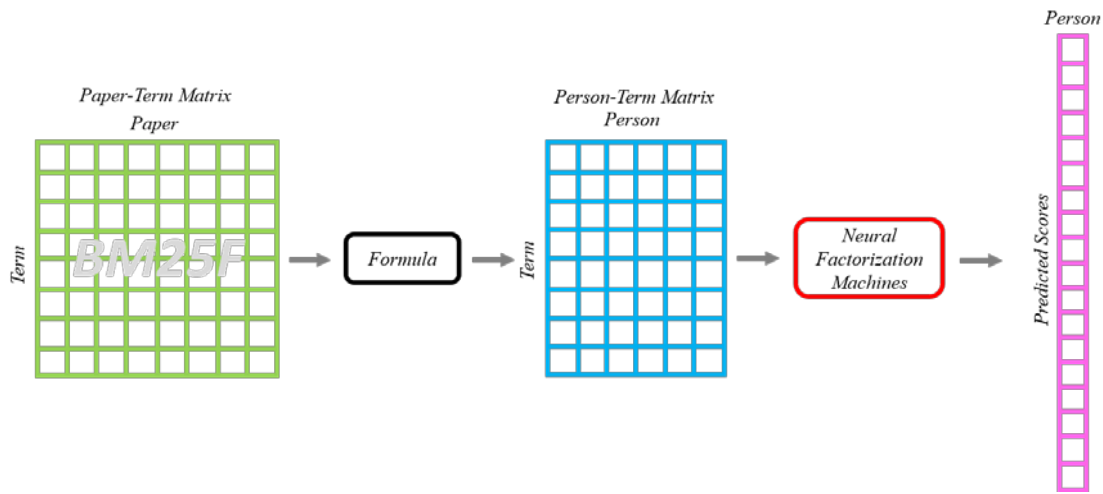

Figure 7. Procedures for searching using Neural Factorization Machines

In order to use the Neural Factorization Machines algorithm, we need to first transform the person-term matrix into Libfm format. The example below demonstrates what should the Libfm format look like for our data:

Example:

Author with ID #2728 wrote term with ID #236991 "deep learning" with score 2.271

Author with ID #2694 wrote term with ID #274922 "machine translation" with score 5.357

Then the Libfm should contain these two lines to represent the examples above:

 2.271 2728:1 236991:1

 5.357 2694:1 274922:1

Before training, we need to split the dataset into training, validation, and testing set to prevent overfitting and to evaluate our model. When the training process completes, we can then use the trained model to predict the scores for researchers given a search query. For instance, when a user searches "Natural Language Processing", assume the term id is #237000, then we can simply generate a prediction Libfm file with the following lines:

Prediction Libfm file:

    0  researcherID0:1  237000:1

    0  researcherID1:1  237000:1

    ……

    0  researcherID20000:1  237000:1

Then we need to feed these data into the trained model to get the score prediction that each researcher will have on this term, then rank researchers by descending order of the predicted score to get the query results.

### 3.1.2 Searching using MAG knowledge base

Another version of the approach we researched and implemented is to do the searching using a hybrid approach, i.e., by combining the traditional keyword matching with Microsoft Academic Graph (MAG) knowledge base. As inspired by the literature reviews, combining domain knowledge when doing searching and recommendation can produce better results that using conventional methods alone cannot achieve. Also, according to MAG publication and its official documentation, it contains a huge amount of knowledge that is suitable to be used in our project, for instance, the Field-of-studies (FOS) tags associated with each publication and the FOS hierarchy structure built using a combination of machine learning, graph analysis algorithms, Wikipedia data, as well as human effort.

The diagram below illustrates the procedures take in this approach.

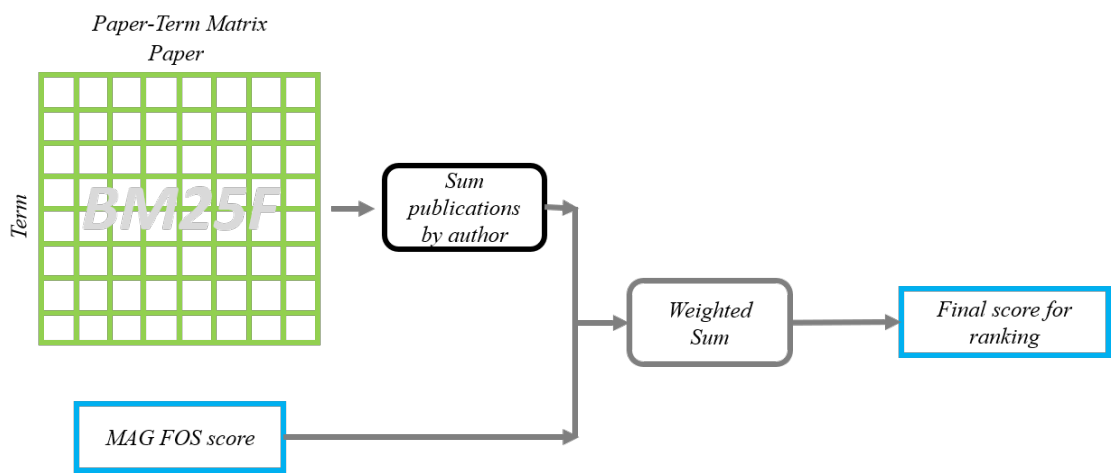

Figure 8. Searching using the hybrid approach

During the implementation, firstly, a paper-term matrix is built using the same way we did in the previous solution's implementation. This matrix contains the BM25F scores for each term in each publication, which measures the degree of relevancy. Secondly, the publication FOS confidence scores are retrieved from MAG, example format of the data is as follows:

Publication ID #297774   FOS ID #6996 (Computer Science): 0.98   FOS ID #3665 (NLP): 0.83   FOS ID #2764 (Machine Translation): 0.69

, which indicates that publication #297774 is related to computer science with a score of 0.98, and is also related to natural language processing with a score of 0.83, etc. This information can be treated as the "product features" for each publication, the same way when we recommend other products to users using recommendation systems techniques. When summing all the publications written by an author, it becomes the features that can be used to characterize that author, and hence, can be used to rank authors, and that is the reason why this knowledge base is useful in this project.

After the matching score and the knowledge base score has been calculated, we take the weighted sum of them to yield the final score, which will be used for ranking the search results, as well as ranking the browsing results.

## 3.2 Browsing
### 3.2.1 Browsing overview

By plan, we carried out researches not only on searching by queries, but also on browsing by academic discipline areas. Basically, we build a concept hierarchy tree, which is a fixed tree, then place researchers under different leaf classes. In each leaf class, researchers are ranked according to the relevancy, in descending order.

Ideally, researchers should be placed under leaf classes base on the following criteria:

1) Each researcher should only appear in a limited number of classes. In our project, a threshold of 7 has been applied to prevent a researcher from appearing in lots of classes.

2) The researchers should only be put into a certain leaf class when his/her relevancy score is higher than a certain threshold. In another word, we do not want to put a researcher who is not so relevant into a leaf class.

3) Every researcher should be placed in at least one leaf class. This criterion is imposed to make sure that all researchers are searchable by students and staffs.

4) If a leaf class has no researcher in it, then it will be removed and will not be displayed.

Browsing can help users to browse and discover the research discipline areas they are interested in, and then find experts under each research discipline areas.

### 3.2.2 Concept hierarchy tree construction

The concept hierarchy tree is built using a combination of data sources: Wikipedia Outline of Academic Disciplines and Australian & New Zealand Standard Research Classification (Field of Research, FoR), published by the Australian Government Research Council. In addition, human inspection effort is used to make sure the combination produces a concept hierarchy tree that totally makes sense.

The diagram below illustrates part of the concept hierarchy tree. Since the concept hierarchy tree is pretty large, the diagram only demonstrates part of the concept hierarchy tree we constructed.

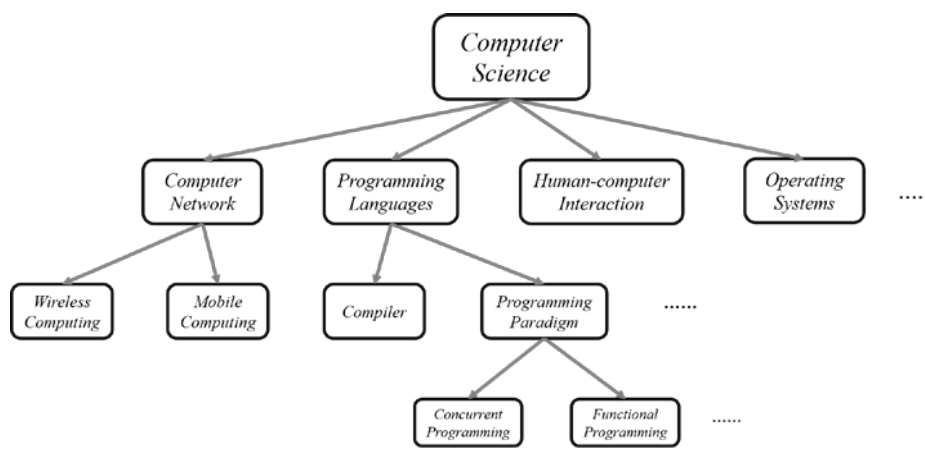

Figure 9. An illustrative example of the concept hierarchy tree

In the concept hierarchy tree, computer network, programming languages, human-computer interactions, etc. are all sub-concepts under computer science, and compilers, programming paradigm, etc. are sub-concepts under programming languages. Researchers will then be classified into different leaf classes.

### 3.2.3 Classification of researchers

After the concept hierarchy tree is constructed, researchers are classified into different leaf classes. The scores used during the classification are the score calculated during searching approach 2 mentioned in previous sections, i.e., calculated using the combination of keyword matching scores and knowledge base scores.

Finally, the concept hierarchy browsing tree is outputted into JSON format and will be deployed into the academic searching platform.

## 3.3 Term extraction
### 3.3.1 Data source used

The term extraction process utilizes two data sources, the Wikipedia pages-articles dump database, and the Microsoft Academic Graph Field of studies. During the implementation of searching approach 1, only the Wikipedia data is used to extract terms, while the implementation of searching approach 2, as well as the browsing, uses both data sources to perform term extraction procedures.

### 3.3.2 Term dictionary construction

For the Wikipedia dataset, it is a noisy dataset, which means some cleansing processes should be performed before constructing the term dictionary. Below are some examples of noisy Wikipedia terms:

    Noisy term 1: File: language.jpg

    Noisy term 2: List of languages by number of native speakers

    Noisy term 3: Computer Science (Outline)

Hence, the following cleansing procedures are performed prior to constructing the term dictionary:

1) All terms that contain Wikipedia Namespaces (File: , User: , Gadget: etc. ) are removed.

2) For these terms that contain parentheses, like (Outline) and (Disambiguation), etc., the parentheses part will be removed, hence, the noisy term 3 will become Computer Science.

3) After step 2, all terms that contain more than 3 tokens will be removed. In this case, we only consider the terms that have at most 3 tokens, due to the observation that most terms contain no more than 3 tokens.

After performing the cleansing, there remain 10048143 Wikipedia terms.

Unlike the Wikipedia data, MAG data is already pre-processed and manually inspected, hence, we can directly use all the Field of Studies terms.

As described above, in the scenario where only Wikipedia data is used, the term dictionary contains all terms from cleaned Wikipedia data, and in scenarios where both Wikipedia data and MAG data are used, the term dictionary contains the union of terms from these two sources.

### 3.3.3 Term extraction mechanism

No matter whether it is a search query entered by the user, titles read from the publication database, or abstract content of the papers, before term extraction, it will be pre-processed using lowercasing, removing duplicate white spaces and punctuations, removing copyright information, performing sentence segmentation, then performing tokenization.

After pre-processing is done, we will extract terms from each sentence. When handling each sentence, we use a context window of length 3, as illustrated in the figure below, each yellow line indicates a context window. If token in the context window is a term in term dictionary, then we move the context window to the end of this term, otherwise, we shrink the context window length by one, and keep repeating. When the length of the context window reaches 1, and we still haven't recognised it as a term yet, then if it is a stop word, we move the context window, otherwise, we lemmatize it and then add it into the result. We keep repeating the same process until we reach the end of the sentence.

The diagrams below illustrate examples of the term extraction process. It demonstrates the concept of context window we used during the term extraction process, how the terms are recognized, how stop words are removed, as well as when not being recognized as an existing term in term dictionary, how a single token term is handled.

```
machine translation researcher loves natural language processing.
machine translation researcher - (not a dictionary term)
machine translation - (recognized as a dictionary term, move context window)
            researcher loves natural - (not a dictionary term)
            researcher loves - (not a dictionary term)
            researcher - (recognized as a dictionary term, then move context window)
                      loves natural language - (not a dictionary term)
                      loves natural - (not a dictionary term)
                      loves - (lemmatize, move context window)
                              natural language processing - (is a dictionary term, move context window)
                              End of the sentence, term extraction completed.

Terms extracted: machine translation, researcher, love, natural language processing
```

Figure 10. Illustrative example I of term extraction process

```
machine translation is a subfield of natural language processing.
machine translation is - (not a dictionary term)
machine translation - (recognized as a dictionary term, move context window)
            is a subfield - (not a dictionary term)
            is a - (not a dictionary term)
            is - (stop word, move context window)
                 a subfield of - (not a dictionary term)
                 a subfield - (not a dictionary term)
                 a - (stop word, move context window)
                     ……
                     natural language processing - (is a dictionary term, move context window)
                     End of the sentence, term extraction completed.

Terms extracted: machine translation, subfield, natural language processing
```

Figure 11. Illustrative example II of term extraction process

Also note that if both data sources have been used, then during term extraction, we give priority to recognize the MAG terms first, if not recognized, then we will start matching Wikipedia terms.

### 3.3.3.1 Advantages of term extraction mechanism
By using this method, we can extract terms from texts base on the term dictionary we constructed. This method enjoys the following advantages:

1) It can extract terms efficiently
2) It's immune to stop words

3) It is immune to derivational morphologies, such as plays – playing – play, or recommended – recommend, etc.

## 3.4 Query auto-completion
### 3.4.1 Data source used

Since in the final deployment, the solution version 2 is used, hence, the query auto-completion uses both MAG data and Wikipedia data. Same as the term extraction, the data has been cleaned and a combined term dictionary is then constructed to be used during query auto-completion.

### 3.4.2 Trie data structure

The prefix trie data structure is constructed for query auto-completion. The reason why prefix trie is used as the main data structure is that it has the following advantages:

1) Insertion and string finding in trie can be done in $O(L)$ time, where $L$ represents the length of the string. This is faster than data structures like BST, it is also better than hashing, since it doesn't require computing any hash function, and doesn't need to handle collisions either.
2) Performing a prefix search is very efficient on a Prefix Trie.

Considering that the query auto-completion should be fast without delays and long waiting times, and it is also a prefix search problem, therefore, prefix trie data structure becomes the best candidate to be used.

### 3.4.3 Prefix trie construction

After constructing the term dictionary, for each of the terms in the term dictionary, we insert it into the prefix trie, and we also store the frequency of each term, i.e., number of occurrences in all publications, for MAG terms only, as we want to do so in order to give priority to MAG terms and let them be displayed at top of the auto-completion suggestions. Finally, the constructed prefix trie is dumped into a dump file for deployment.

### 3.4.4 Query auto-completion suggestions

Using the constructed prefix trie, when the user is typing a partial search query, we can return corresponding strings in the prefix trie which matches the partial search query, and all the suggestions are ranked in descending order of occurrence frequency pre-stored in the prefix trie. Note that we start giving query auto-complete suggestions when the user has typed at least three characters.

## 3.5 Server connection

In order to put the platform that we implemented into reality, a TCP client and server is developed to support the searching service. A consistent TCP server, which is capable of handling multiple requests simultaneously, takes user's search query in string format and then returns the search result back to the client. A consistent TCP client is also developed, which is capable of sending searching requests to the server, then waiting and receiving the query results. After receiving the query result, instead of terminating immediately, the TCP client will still maintain a long-term connection with the server. This can make sure that the following requests will not face the delay caused by establishing a connection with the server again and again, which dramatically improves the searching efficiency.

The client and server are not only efficient, but also reliable. By using TCP to establish the client-server connection, it builds a reliable communication channel between two sides, which will make the suffering of network delays and package lost no longer an issue for our system.

# 4. Data collection and analysis

## 4.1 Publication data from FindExperts

### 4.1.1 Dataset overview

In this project, we download the publication dataset from the University of Melbourne FindExperts platform. The dataset contains 398254 publications with 19 columns, among all these 19 columns, the following is used by our project.

Columns used in publication dataset

| Title | Publication title |
|---|---|
| Abstract | Publication abstract |
| Journal Name | The name of the journal in which the publication is published |
| Conference Name | The name of the conference in which the publication is accepted |
| Claimed Users | Name of authors of publications |
| Keywords | Publication keywords |

Table 1. Columns used in publication dataset

### 4.1.2 Dataset cleansing and pre-processing

As described in the above section, before using the data in this dataset, we performed the following pre-processing procedures:

1) Lowercasing
2) Removing copyright information such as © 2013 Springer
3) Sentence segmentation, which segments paragraphs into sentences
4) Tokenization, which tokenizes sentence into tokens

An illustrative example of pre-processing steps and corresponding outcomes is shown in figure 3 in the section 3.1.1.2 Publication data pre-processing.

### 4.1.3 Dataset usage

During this project, the publication dataset is used in the following scenarios:

1) Searching approach 1, used as the data source to measure the relevancy between the query term and researchers.

2) Searching approach 2, used as the data source to calculate the BM25F score of the query term for each researcher, then combined with scores from the MAG knowledge base to perform ranking.

3) Browsing, used as the data source to calculate the BM25F score of leaf classes for each researcher, then combined with scores from the MAG knowledge base to perform the ranking.

4) Query auto-completion, used to provide term occurrence frequencies when constructing the term dictionary.

## 4.2 Wikipedia dumps

### 4.2.1 Dataset overview

The Wikipedia dump file is downloaded from Wikimedia dump service enwiki. Before cleansing and pre-processing, it contains 19549637 term entries, after cleansing and pre-processing, there remain 10048143 terms.

### 4.2.2 Dataset cleansing and pre-processing

As mentioned in the above sections, the Wikipedia dump file is a little bit noisy, hence, before using it, we removed Wikipedia Namespaces entries such as "File: language.jpg", which denotes a picture file in Wikipedia, but not useful for our project. Secondly, Wikipedia includes some short annotations like "(Outline)" or "(Disambiguation)" for easy browsing, we removed these short annotations. Finally, given the observation that most terms contain at most three tokens, therefore, we removed all entries which have more than three tokens. Thus, terms like "List of languages by number of native speakers", which is obviously not a valid term, are eliminated from the dataset.

### 4.2.3 Dataset usage

During this project, the Wikipedia terms dataset is used in the following tasks:

1) Searching approach 1 and 2, used to build the matrices
2) Construction of the term dictionary, used for term extraction and query auto-completion

## 4.3 Wikipedia category hierarchy construction

Initially, we constructed a Wikipedia category hierarchy. The construction is done using the Wikipedia API, the details of the API is as follows:

```
wikipedia_api_url = https://en.wikipedia.org/w/api.php
parameters = {
    'action': "query",
    'list': "categorymembers",
    'cmtitle': category,
    'cmlimit': 100000,
    'format': "json"
}
```

By using this API, we can easily retrieve the subcategories and pages affiliated to any Wikipedia categories.

The reason why we decided to construct such a category hierarchy initially is that we planned to take parent categories into account when calculating the scores in the searching service. For instance, term natural language processing is the parent category of machine translation, and term computer science is the parent category of natural language processing. When we see the term machine translation happens in a researcher's publication, then we also added some scores to the natural language processing and computer science for this researcher. By doing so, when a user searches for researchers who are related to computer science, then this researcher becomes possible to be searched. Even though he/she does not mention natural language processing or computer science in his/her publications explicitly, by taking this category hierarchy into account, we can still use the hierarchy to trace towards the top hierarchy and reward this researcher a score on term computer science. However, in our final implementation, this idea is currently not used. The reasons are as follows:

1) The category hierarchy constructed using Wikipedia is very noisy, an example of Wikipedia category hierarchy is illustrated in the table below.
2) We later discovered and utilized MAG knowledge base, which also has a term hierarchy built in it, it's also based on Wikipedia data, the differences are it applied advanced machine learning algorithms, as well as graph link analysis algorithms to perform concept discovery and thus refine the term hierarchy, it also applied human inspection to make sure the hierarchy built is in excellent quality.

3) In addition to the concept hierarchy, the MAG knowledge base also has FOS tags associated with each publication with a score, so the idea of counting parent categories is no longer necessary, as the FOS tags contained in the MAG knowledge base already implemented this idea for us. If a paper is related to machine translation, it will also be tagged with natural language processing and computer science.

The tables below compare the child concepts under the term computer science for both the Wikipedia category hierarchy and MAG concept hierarchy. Apparently, the Wikipedia category hierarchy is noisier than MAG concept hierarchy.

| MAG data: Children concepts under term "Computer Science" | | | |
|---|---|---|---|
| Computational Science | Computer Hardware | Algorithm | Information Retrieval |
| Speech Recognition | Computer Network | Computer Vision | Computer Security |
| Simulation | Multimedia | Knowledge Management | Telecommunications |
| Database | Real-time Computing | Theoretical Computer Science | Human-computer Interaction |
| Internet Privacy | Operating System | Computer Engineering | Software Engineering |
| Computer Architecture | Machine Learning | Distributed Computing | Computer Graphics |
| Data Mining | World Wide Web | Embedded System | Artificial Intelligence |
| Library Science | Parallel Computing | Pattern Recognition | Programming Language |
| Natural Language Processing | Data Science | | |

Table 2. Children concepts under the term "Computer Science" in MAG data

| Wikipdia data: Children concepts under term "Computer Science" | | | |
|---|---|---|---|
| Glossary of Computer Science | Boolean | Clever Score | Computer Engineering |
| Outline of Computer Science | Trace Cache | K-d heap | Visual Computing |
| Software | Computer Science in Sports | Programmer | …… |

Table 3. Children concepts under the term "Computer Science" in Wikipedia data

Considering these reasons, we did not use the Wikipedia category hierarchy we constructed in our final implementation, the publication associated with FOS tags in MAG data has already implemented this idea for us.

## 4.4 Microsoft Academic Graph data
### 4.4.1 Data overview

Inspired by our literature reviews, we decided to integrate Microsoft Academic Graph knowledge base into our project. The Microsoft Academic Graph (MAG) knowledge base is the largest academic knowledge base to-date, it contains 3 components as follows:

1) 664845 Field-of-Study concept terms identified using Wikipedia data, machine learning algorithms like graph link analysis, entity type based filtering and enrichment, as well as human inspections.
2) Publications tagged with concepts and scores associated with each tagging. Done by formulated as a multi-class classification problem.
3) A 6 levels concept hierarchy.

All data are already useable, there are no pre-processing steps required.

The table below illustrates the data tables contained in Microsoft Academic Graph (MAG) knowledge base.

| Microsoft Academic Graph data tables | | | |
|---|---|---|---|
| Affiliation | Author | FieldsOfStudy | FieldsOfStudyChildren |
| RelatedFieldsOfStudy | PaperFieldsOfStudy | Papers | PaperAuthorAffiliations |
| ConferenceInstances | Journals | PaperAbstractInvertedIndex | PaperCitationContext |
| PaperResources | PaperUrls | FieldOfStudyExtendedAttributes | ConferenceSeries |
| PaperReferences | PaperRecommendations | EntityRelatedEntities | PaperExtendedAttributes |

Table 4. Microsoft Academic Graph knowledge base data tables

The table below demonstrates which data tables are useful in this project, the description of the content they contain, as well as the data schema.

| MAG data tables used in this project | |
|---|---|
| Affiliation | Information about affiliations<br>PRIMARY KEY(AffiliationId (long)),<br>DisplayName (string) |
| Author | Information about authors<br>PRIMARY KEY(AuthorId (long)),<br>DisplayName (string) |
| FieldsOfStudy | Information about Fields-of-Studies<br>PRIMARY KEY(FieldsOfStudyId (long)),<br>DisplayName(string) |
| FieldsOfStudyChildren | Information about children of each Field-of-Study, if there are any<br>FOREIGN KEY(FieldsOfStudyId (long)) REFERENCES FieldsOfStudy(FieldsOfStudyId),<br>FOREIGN KEY(ChildFieldOfStudyId (long)) REFERENCES FieldsOfStudy(FieldsOfStudyId) |
| RelatedFieldsOfStudy | Information about the related Field-of-Studies of each Field-of-Study, if there are any<br>FOREIGN KEY(FieldOfStudyId1 (long)) REFERENCES FieldsOfStudy(FieldsOfStudyId),<br>FOREIGN KEY(FieldOfStudyId2 (long)) REFERENCES FieldsOfStudy(FieldsOfStudyId) |
| PaperFieldsOfStudy | Field-of-Studies associated with each paper, and the corresponding score<br>FOREIGN KEY(PaperId (long)) REFERENCES Papers(PaperId),<br>FOREIGN KEY(FieldOfStudyId (long)) REFERENCES FieldsOfStudy(FieldsOfStudyId),<br>Score (float) |
| Papers | Information about papers<br>PRIMARY KEY(PaperId (long)),<br>PaperTitle (string), BookTitle (string), Year (int), Publisher (string),<br>FOREIGN KEY(JournalId (long)) REFERENCES Journal(JournalId),<br>FOREIGN KEY(ConferenceInstanceId (long)) REFERENCES ConferenceInstances(ConferenceInstanceId),<br>Volume (string), Issue (string),<br>FirstPage (string), LastPage (string) |
| PaperAuthorAffiliations | Authors affiliated with each paper<br>FOREIGN KEY(PaperId (long)) REFERENCES Papers(PaperId),<br>FOREIGN KEY(AuthorId (long)) REFERENCES Author(AuthorId),<br>FOREIGN KEY(AffiliationId (long)) REFERENCES Affiliation(AffiliationId) |
| ConferenceInstances | Information about conferences<br>PRIMARY KEY(ConferenceInstanceId (long)),<br>DisplayName (string) |
| Journals | Information about journals<br>PRIMARY KEY(JournalId (long)),<br>DisplayName (string) |

Table 5. MAG data tables used in this project

### 4.4.2 Data usage
During this project, we use the MAG knowledge base in the following cases:

1) Searching, in which knowledge base scores are used for query result ranking.
2) Browsing, in which knowledge base scores are used to decide which leaf classes should a researcher belongs to, as well as ranking researchers in each leaf class.
3) Term extraction, in which MAG FOS terms are used to build the term dictionary.
4) Query auto-complete, in which MAG FOS terms are used to build the term dictionary.

# 5. Deployment, results, and findings

## 5.1 Result comparisons between two searching approaches

For the searching functionality, two approaches have been implemented as we described above, and their results are compared thoroughly to see which approach performs better. During result comparison, several different search queries in several discipline areas including computer science, biology, etc., have been compared. Below I will demonstrate examples of the result comparison.

Below is the search result with the query "Natural Language Processing".

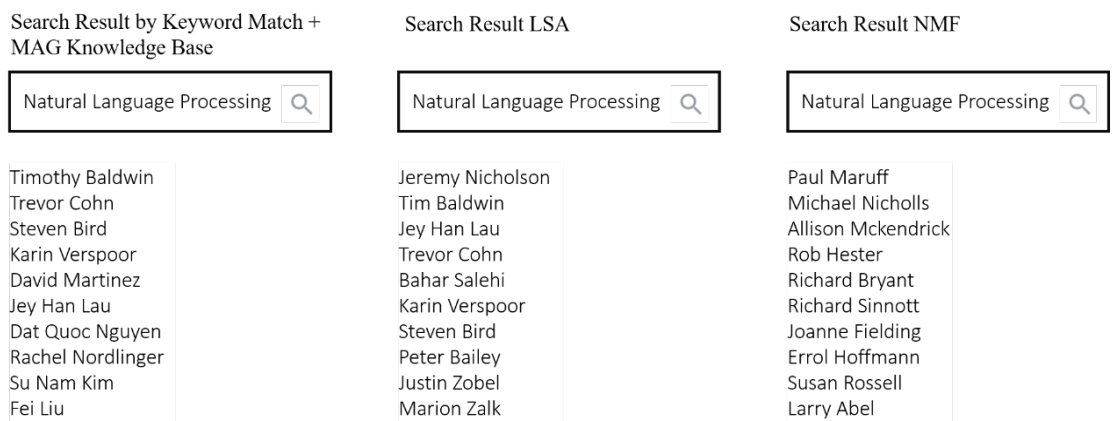

Figure 12. Comparison of search results of the hybrid approach, LSA and NMF

Below is the search result with the query "Data Mining".

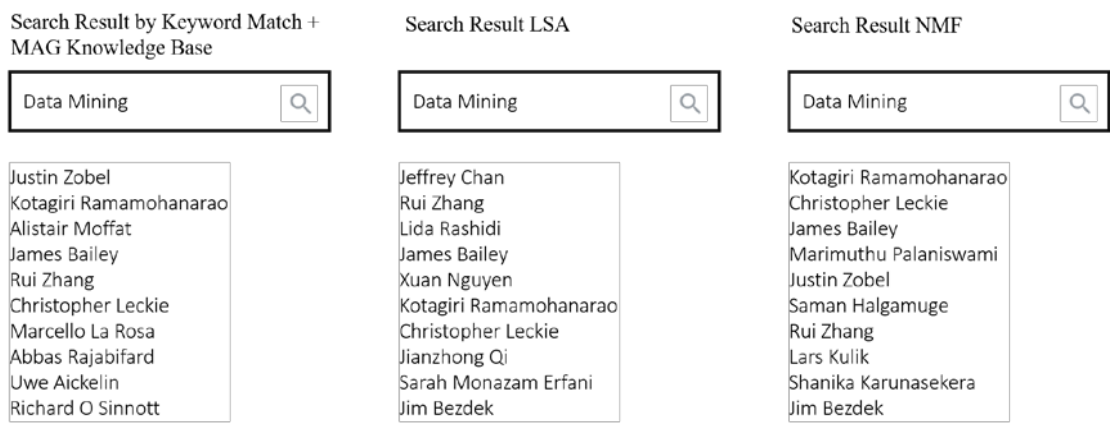

Figure 13. Comparison of search results of the hybrid approach, LSA and NMF

These two query results are just two examples among various search queries we experimented with. As we can see, for both of them, the search result returned by keyword matching with MAG knowledge base makes sense, the researchers who are relevant to this research area is ranked highly, the ranking is in the descending order of relevancy. However, for the results produced by LSA, although the researchers in the result are relevant to the search query, the order in which the researchers are ranked is sometimes incorrect. The researcher who is highly relevant to the query is not ranked in top place, and this is the case for some other queries we tried as well. Hence, compared with search results produced by keyword matching plus MAG knowledge base, the search result produced by LSA is inferior to the former. As for the result produced by NMF, there are some neuroscientists and behavioural scientists who have neural and language happening frequently in their papers get ranked highly in the result of NLP, while we cannot say that they are strictly irrelevant to NLP, but compared with researchers in the school of computer science, they are inferior in the research area of NLP. Hence, they should not be ranked in top places, although they are relevant. The result returned by NMF also suffers from the ranking order issue confronted by LSA as well. Hence it turns out that neither LSA nor NMF performs better than searching using keyword matching plus MAG knowledge base.

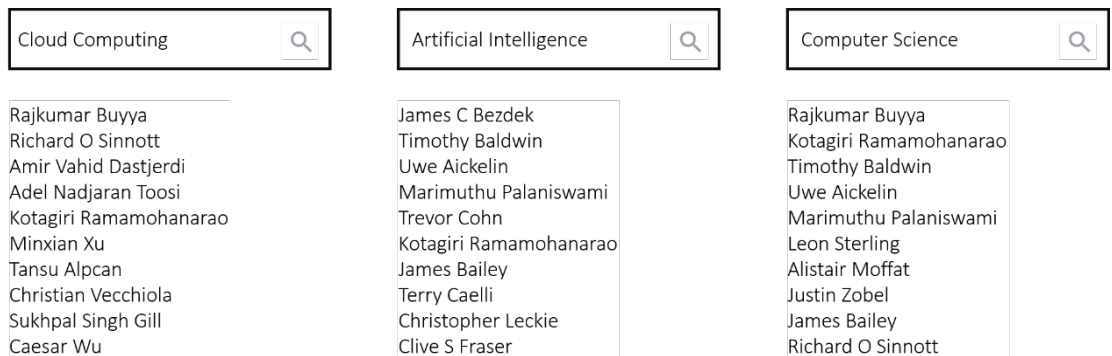

Figure 14. Demonstration of search results of the hybrid approach

The above figure shows three more result examples produced by our hybrid approach. Experimental results demonstrate that the hybrid approach produces results that meet our expectations.

### 5.1.1 Knowledge gained
The search results we got clearly demonstrates the superiority of the hybrid approach compared with both LSA and NMF, which indicates that combining the domain knowledge can indeed help in this case. Also, under this scenario in which we don't have enough ground truth ranking data, the domain knowledge contained in the knowledge base can sort of acting as the ground truth, which explains why under this scenario, the hybrid approach performs the best. Hence, when facing a similar scenario in the future, combining domain knowledge from a knowledge base is a sound approach to try.

## 5.2 Browsing results demonstration

Given that among all the variations of searching implementation we experimented, the combination of MAG knowledge base and keyword matching yields best performance, hence, we used this method during our browsing implementation.

Below is a demonstration of part of the concept hierarchy tree structure and corresponding researchers in each leaf class.

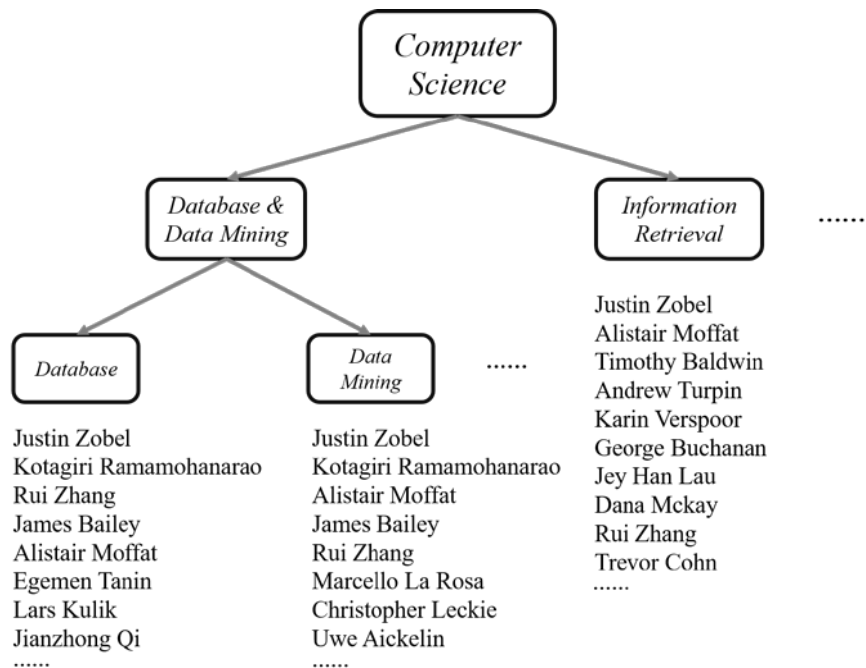

Figure 15. Demonstration of browsing results

As we can see, in this concept hierarchy tree, highly relevant researchers are categorized into corresponding leaf classes correctly, in the descending relevancy order that makes sense, hence, browsing using our searching and ranking method works pretty well.

### 5.2.1 Discussions of browsing criteria

Initially, we only applied one criterion for browsing to satisfy, which is the qualifying threshold. The result returned under that case is a little bit unsatisfactory due to the following reasons:

1) Some researchers appear in lots of leaf classes, while some researchers don't appear in any leaf classes at all.
2) Although some Fields-of-study of some researchers meets the qualifying threshold, this FOS is not the major research interest of this researcher, hence, satisfying this qualifying threshold does not indicate that this research area is the major research interest of this researcher.

After observing these issues, we updated our browsing criteria. Firstly, if a researcher doesn't have any FOS that can satisfy the qualifying threshold, then we categorize this researcher into his/her top-ranked research area, even if its score doesn't meet the qualifying threshold. Also, instead of only using a qualifying threshold, we also applied a total number of qualification threshold, which is set to 7. Hence, for each researcher, he/she can only enter $C$ leaf classes, where $C$ is defined as follows,

$$C = \min(7, \#\ FOS\ for\ this\ researcher\ which\ meets\ qualifying\ threshold)$$

After updating the browsing criteria, the browsing results become better and meet our expectations.

### 5.3 Query auto-complete result demonstration

After thorough experiments, our query auto-complete mechanism also performs well. Below is an example of auto-complete suggestions when the user is typing "machine learning".

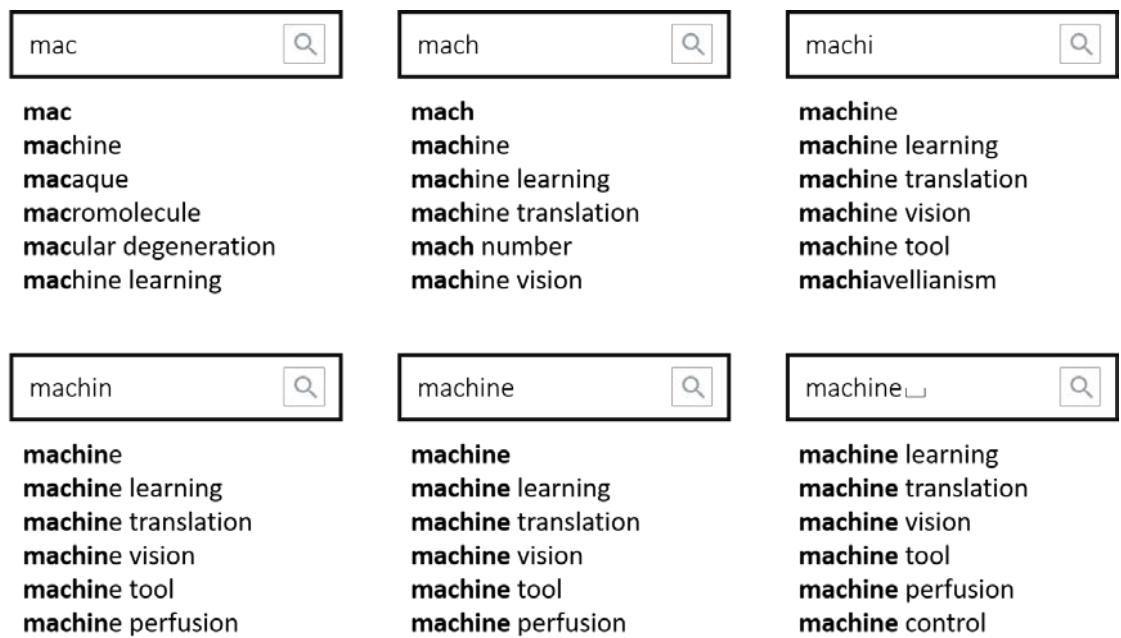

Figure 16. Demonstration of query auto-complete suggestions

During the typing process, auto-complete suggestion entries are displayed based on the frequency they occur in the MAG knowledge base plus our publication dataset. As we can see, when the user types "machin", frequent terms like machine, machine learning, machine translation, machine vision, etc. are displayed in top places to help the user complete his/her search query. The query auto-complete performs well and meets our expectations.

# 6. Discussion and analysis

## 6.1 Normalization of NFM input in searching

Based on the experimental results, we think that firstly, the reason why Neural Factorization machine does not perform well is that this algorithm is usually used on scaled data, for instance, movie rating data will usually sit in the range between 0 and 10, mobile app download data, which is binary, the user will choose to either download, or not download the app. However, in our case, the input data is the BM25F scores extracted from publications. If a researcher publishes lots of papers that frequently mention a term, then the BM25F score of this term for this researcher will be very high, it can even reach 1000 by our experiment. Hence, unlike the usual datasets that this algorithm works on, our dataset is not scaled. After realizing this, we applied normalization on the scores before feeding them into the algorithm, unfortunately, the normalized data doesn't seem to produce satisfactory ranking results. We will keep investigating the form of input data we should use for this algorithm in our future works.

## 6.2 Hyperparameter tuning in searching

Secondly, the matrix transformation formula we proposed contains some hyperparameters that we need to be tuned, the tuning results of these hyperparameters will directly affect the results produced by LSA, NMF or NFM algorithm. However, as we mentioned above, this project faces a huge challenge of lacking ground truth ranking data, hence, making traditional hyperparameters tuning methods like grid search infeasible. During the project, hyperparameter tuning relies on manual effort, i.e., manually adjust the hyperparameter settings, then manually inspect various search query results to see whether they are in the correct ranking as we expected. Hence, the process is both time and human effort intensive. As we can see, the results produced by LSA and NMF generally make sense, however, the ranking of some researchers are still not perfect. Hence, finer hyperparameter tuning should be performed as future works, but generally speaking, given that the hyperparameter is not perfectly tuned, both LSA and NMF works reasonably well.

## 6.3 Discussion of searching using the hybrid approach

The MAG knowledge base plus keyword matching approach can avoid issues we mentioned above perfectly. Firstly, the keyword matching we used in this solution doesn't require transforming the Paper-term matrix into Person-term matrix, hence, the hyperparameter tuning issue can be avoided. Secondly, this method does not have any requirements about the scale of the input, as it calculates the score using a weighted sum of scores. More importantly, this method utilizes the domain knowledge contained in the MAG knowledge base, which can help us to refine the search results and produce search results with higher quality. Hence, this method is the one we used finally in our searching and browsing.

## 6.4 Discussion of browsing

By utilizing the hybrid approach, the browsing performs perfectly as we expected. Under each leaf class, highly relevant researchers are included and ranked in the correct order as we expected. This demonstrates again the effectiveness of the hybrid approach we experimented with in searching. What's more, the updated browsing criteria we discussed above also works fine, the results produced under updated browsing criteria don't suffer from noises anymore.

## 6.5 Discussion of query auto-completion

Query auto-complete also works well, its query suggestion results are also satisfactory. Popular query suggestions will be provided to users to help them complete their search queries.

Another idea we planned but haven't put into reality is query correction. Our previous implementation utilizes edit distance, which is a common approach to perform query correction. However, in order to make it efficient online, we need to perform all computations beforehand, which may potentially require a long pre-computation time, as well as huge storage spaces. We will keep investigating how to perform query correction efficiently in our future works.

# 7. Conclusions and contributions

In this project, we experimented different methods to perform researcher searching, the method we implemented and experimented including using Latent Semantic Analysis, with the person-term matrix as input, using Non-negative Matrix Factorization, with the person-term matrix as input, using Neural Factorization Machine, a deep learning version of the Factorization Machine, with the person-term matrix as input, and a hybrid approach, which combines the score from Microsoft Academic Graph knowledge base, and the score we calculated based on keyword matching. After thorough experiments and analyses, we choose to use the hybrid approach as our final way of performing searching and ranking, as it does not face the issues like hyperparameter tuning, etc., and experimentally, it produces the best query search result, compared with other methods we experimented.

Besides searching, we also performed browsing researchers by discipline areas, the ranking mechanism we used in browsing is the same as we used during searching.

In addition to searching and browsing, other functionalities we investigated and implemented including query auto-complete, field-of-study information displaying, as well as constructing a server and client to make the searching functionality deployable.

The work of this project truly enhances the academic searching experience of both students and staffs, and also contributes towards academic searching and browsing. The project investigated several algorithms and made thorough comparisons, we also combined the academic knowledge base to enhance the searching and browsing results. More importantly, we experimented on how to produce satisfactory searching and browsing results given a limited amount of ground truth data, and hence, similar scenarios people encountered later can be handled in similar ways.

# 8. Future works

Future works of this project including the following:

1) When calculating the score of each researcher, instead of all his/her research and publications equally, we can apply weights by the time the research is done, or the publication is published. Since firstly, the research interest of a researcher can change gradually as time passes, the research he/she carried out 20 years ago should be weighted far less than the research he/she conducted a month ago. And secondly, the depth of the research can change dramatically over time, for instance, the research regarding quantum computing 20 years ago is absolutely different from current quantum computing researches. Hence, different weights should be applied with regard to time.

2) Our platform also allows the researcher to enter the research interest by itself, as well as projects that he/she is currently carrying out. Hence, when performing searching, we can also take user input research interests, as well as project information into account.

3) More effort should be spent to make the methods we used capable of handling of periodic updates of publications, research areas, etc.